# The application of Graphene as a sample support in Transmission Electron Microscopy

Radosav S. Pantelic[a], Jannik C. Meyer[b], Ute Kaiser[c]*, Henning Stahlberg[a]*

(a) Center for Cellular Imaging and Nanoanalytics, Biozentrum, University of Basel, Basel, Switzerland

(b) University of Vienna, Department of Physics, Vienna, Austria

(c) Electron Microscopy Department of Materials Sciences, University of Ulm, Ulm, Germany

(*) To whom correspondence should be addressed.

**Abstract**

Transmission electron microscopy has witnessed rampant development and surging point resolution over the past few years. The improved imaging performance of modern electron microscopes shifts the bottleneck for image contrast and resolution to sample preparation. Hence, it is increasingly being realized that the full potential of electron microscopy will only be realized with the optimization of current sample preparation techniques. Perhaps the most recognized issues are background signal and noise contributed by sample supports, sample charging and instability. Graphene provides supports of single atom thickness, extreme physical stability, periodic structure, and ballistic electrical conductivity. As an increasing number of applications adapting graphene to their benefit emerge, we discuss the unique capabilities afforded by the use of graphene as a sample support for electron microscopy.

**Introduction**

High-resolution transmission electron microscopy (HR-TEM) and scanning transmission electron microscopy (STEM) of thin samples provides rich and versatile information at unsurpassed spatial resolutions, often difficult if not impossible to obtain by any other technique. Combining these techniques with other technologies such as electron energy loss spectroscopy and energy-dispersive x-ray analysis provides further insight into the compositional and functional characteristics of samples. Transmission electron microscopy is also used to investigate the 3D organization of biological macromolecules and assemblies, bridging the sample gap left by other techniques such as Nuclear Magnetic Resonance spectroscopy and X-Ray crystallography. When combined with experimental sample conditions and preparation methods, a wealth of structural and functional information can be extrapolated.

Electron microscopy has recently seen a dramatic improvement in instrumentation with the emergence of hardware aberration correction [1-3]. However, imperfections in the preparation of samples often become the limiting factor. Nano-materials, as well as biological molecules are typically prepared across support films that inevitably introduce an additional background signal, attenuating that of the sample. Graphene shows great potential in optimizing the preparation of nano-materials and biological samples in electron microscopy. Graphene layers are of single atom thickness with regular/periodic structure, demonstrate high electrical conductivity and are relatively stable under the electron beam. In this manuscript we discuss and demonstrate some of the outstanding capabilities of graphene TEM supports, the state of the art as well as progress made in applying graphene supports in both biological (especially cryo-electron microscopy) and materials sciences electron microscopy.

**Considerations in Cryo-Electron Microscopy and use of sample supports**

Cryo-electron microscopy (Cryo-EM) refers to the electron microscopy of frozen-hydrated protein complexes, cells and tissue prepared by rapid vitrification [4-6]. This approach is the accepted standard as flash-freezing retains bound water, thus preserving ultrastructural detail in a near-native state. The EM database (EMDB, www.ebi.ac.uk/pdbe/emdb/statistics_num_res.html) catalogues biological structures determined by EM. While modern electron microscopes routinely reach resolutions beyond 2Å, only 20% of deposited structures determined by cryo-EM have reached sub-nanometre resolution, with a mere 3% exceeding 5Å resolution. The stark discrepancy demonstrates the limitation stemming from the delicate nature of biological samples, imposing limitations on handling and imaging. Hence, it is increasingly being realized that an effort to push the resolution of Cryo-EM will also require optimized sample preparation techniques [7-9].

Cryo-EM samples are weak-phase objects, in that electrons passing through vitrified protein as opposed to amorphous ice demonstrate an almost negligible difference in phase shift (*elastic* interaction, ~14 mrad/nm and ~33.84 mrad/nm respectively) [10]. Unstained macromolecular complexes and small proteins consequently demonstrate poor contrast. Cryo-EM samples are also particularly sensitive to radiation damage caused by *inelastic* interaction with the electron beam, necessitating low-dose imaging (i.e. imaging at total electron doses below 20-30 $e^-/Å^2$ at the specimen, or <50-100 $e^-/Å^2$ cumulative per tomogram). This low dose tolerance further reduces the signal to noise ratio (SNR) of acquired data due to the presence of shot noise.

Beam-induced resolution loss also represents a major obstacle to ascertaining high-resolution information from frozen-hydrated samples [11]. Inelastic interaction between the primary electron beam and sample releases Auger and secondary electrons that introduce areas of net positive charge and subsequent charging effects across the sample. This effect is exaggerated when imaging vitrified samples, particularly at liquid helium temperature since the vitreous ice is an insulator [12]. As charge accumulates, images demonstrate a truncation of resolution resembling that of specimen drift. This phenomenon is especially exaggerated upon tilting (i.e. electron crystallography) [11, 13]. A completely alternative model proposes that these effects are not in fact attributed to charging, but rather that radiation damage within the specimen produces stochastically distributed physical stresses across the sample that induce similar sample instability [10]. Charge induced drift has been described as a manifestation of repulsive Coulomb forces exerted between immobile surface charges creating an overall instability across the sample [14]. The phenomenon has also been attributed to a lensing effect, whereby varying electric fields perpendicular to the sample induce image shift when tilted [11, 15].

Several applications in cryo-EM (particularly 2D electron crystallography and imaging at liquid helium temperatures [7, 13]) have demonstrated improved sample stability upon including additional amorphous carbon layers (to presumably dissipate specimen charge or physically stabilize the specimen) [14, 15]. Low concentration samples or preparations requiring multiple steps (e.g. washing to remove unwanted low molecular weight constituents) also necessitate an additional amorphous carbon support to attach and retain protein concentrations [16]. Outside Cryo-EM, thin amorphous carbon has also been reported to reduce charging and improve the stability of plastic embedded tissue and cell sections [14]. In general, amorphous carbon is widely used as a sample support in current and emerging [17, 18] methods in life science TEM.

However, although the inner bulk of amorphous carbon is conductive the surface is electrically insulated and films below ~4 nm in thickness demonstrate almost no electrical conductivity [19].

Only from ~5.6 nm does conductance begin to increase linearly with thickness [14]. However, amorphous carbon is a semiconductor rather than metal, and also suffers significantly reduced conductivity at low temperatures (particularly helium temperatures [12]). Furthermore, when imaged at higher magnification, amorphous carbon supports introduce strong background signal. Consequently, this background signal attenuates and even obscures that of unstained, vitreous samples [15].

**Considerations for materials science TEM and the use of sample supports**

Materials science electron microscopy encompasses a wide range of studies investigating 3- 2- and 1- dimensional volume defects, material interfaces and dopants as well as nano-scale materials such as nanoparticles, nanowires and nanotubes (to name a few) - All of which necessitate atomic resolution. Bulk materials are conventionally thinned down to the required nanometre-scale thickness (< 20 nm, depending on the material and the accelerating voltage used) by grinding, milling, and ion-polishing/milling or other techniques such as focused ion beam (FIB) milling [20]. However, by virtue these samples do not require a support film. Nano-scale objects can often not be prepared as freestanding samples. Our discussion will therefore mainly consider the study of small isolated objects such as nanoparticles, inorganic molecules, nanocrystals, quantum dots and nano-tubes/nanowires. Increasing interest in these novel, nano-scale materials has introduced challenges with sample preparation often unique to each study.

In the past, materials studies have more often been hampered by resolution-limiting electron-optical aberrations. However, with the introduction of aberration-corrected TEM the focus has shifted to those limitations presented by sample preparation. Gold nanoparticles have been prepared across amorphous carbon supports and used as test specimens for assessing instrument performance [21]. Individual heavy atoms can be visualized easily when using ultra-thin amorphous films since beam-induced migration is still slow enough to obtain high-resolution (high-dose) images [22, 23]. However, recent years have seen a tremendous interest in low-dimensional and light-element materials. High-resolution images of single-walled carbon nanotubes (SWCNT's) [24, 25] have become common sight alongside lattice-resolution images of graphene [26, 27]. Moreover, chemical reactions can now be studied inside carbon nanotubes (CNT's) [28]. The resolution and SNR that can be obtained from images are sufficient to detect single-atom vacancies [26, 29], topological defects [25, 27], exact atomic configurations across grain boundaries [30] and amorphous inclusions [31, 32]. However, it should be obvious that imaging at this precision would be difficult if not impossible had an amorphous carbon support been used. Furthermore, sample platforms are of particular importance to dynamic studies and in-situ experiments.

It should be emphasized, that many of these new materials have so far only been imaged in freestanding geometries. Clusters and nano-particles typically cannot be prepared as freestanding samples. Smaller molecules (in particular endohedral fullerenes and metal nanoparticles) have been imaged by HR-TEM after insertion to SWNT's and subsequent preparation as a freestanding sample [33-37]. However, limited space and harsh insertion procedures limit the applicability of SWNT containers in HR-TEM/STEM. The success of this method does however present a strong case for developing similar low-contrast graphene-based supports for a wider variety of samples.

**Graphene and the revival of crystalline TEM supports**

Crystalline supports demonstrate almost no phase contrast down to the resolution of their periodicity regardless of thickness. By reducing support thickness, background amplitude contrast (noise) introduced by secondary and multiple electron scattering within the bulk of the support can also be minimized [38]. Dobelle & Beer (1968) first demonstrated the benefits of crystalline TEM supports in structural biology with the cleavage of graphite [39]. However, its difficulty and limited efficiency ensured eventual obscurity. Similarly, the use of thin graphite supports in material science had already been explored several decades ago, with the visualization of clustered and individual metal atoms bound to the atomic steps of multiple graphitic layers [40].

Graphene has renewed interest in crystalline TEM supports [7, 8, 17, 38, 41-47]. Pristine graphene is essentially electron transparent down to a resolution of 2.13Å, which is still outside resolutions routinely resolved in cryo-EM [38]. In material studies by HR-TEM, the periodic structure of graphene yields a diffracted signal that can be easily Fourier filtered from images as necessary [48]. Furthermore, a single-layer thickness of 0.34 nm (single-atom thickness) [49] contributes only minimum background noise. Yet, the highly ordered structure of graphene is remarkably strong both mechanically and elastically [50-52]. Most interestingly, graphene is a "ballistic" electrical conductor, demonstrating electrical conductivity more than 6 orders of magnitude higher than that of amorphous carbon (converted to bulk units and assuming a thickness of 3.4 Å)[53-55]. High electrical conductivity is also demonstrated at liquid Nitrogen temperatures [56, 57] with charge mobility even increasing at liquid Helium temperatures [58].

**Graphene supports in life sciences TEM**

Deposition of graphene oxide from solution referred to as "drop casting" was a simple and effective way of producing graphene TEM supports [42, 59] and was somewhat reproducible by checking solution concentrations according to UV-VIS absorption spectra [42 ]. Most importantly, the presence of functional groups (carboxyl, hydroxyl and epoxy [60, 61]) rendered supports sufficiently hydrophilic for the application of biological samples and provided a precursor for further functionalization. An interesting example is the decoration of graphene oxide sheets with streptavidin by which biotinylated proteins can be directly purified (by affinity) across highly transparent TEM supports [41].

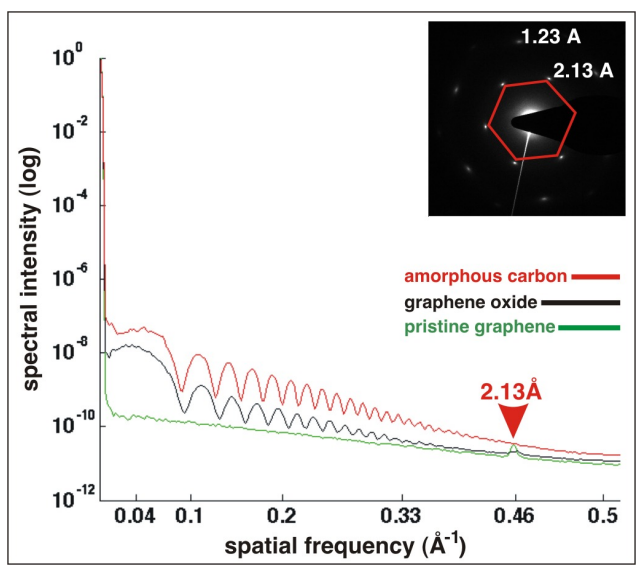

**Figure 1, Comparison of background signal from graphene supports:** Power spectral densities calculated from images (~200 nm defocus, 0.87 Å pixel size) of thin amorphous carbon (red, ~3 nm thickness), monolayer graphene oxide (black, ~1 nm thickness, partially reduced at ~300°C in air [37]) and monolayer pristine graphene (green, ~0.34 nm thickness, rendered pristine at 400°C in vacuum) substrates. Insets show corresponding 2D power spectra for the pristine graphene (A) and graphene oxide (B) samples - note the first diffracted periodicity at 2.13 Å and difference in phase contrast apparent by the appearance (or lack, A) of Thon rings (B).

The predominantly crystalline structure of graphene oxide demonstrates transmission properties approaching those of pristine graphene with sparse, nanometre-scale amorphous defects only contributing weak phase contrast that rapidly tapers off at higher resolution (Fig.1, black). Single layer areas (~1 nm) also exhibit significantly reduced inelastic scattering and therefore reduced background amplitude (noise). However, the random deposition of individual, few-micron sized graphene oxide sheets from solution, produced samples with irregular thickness (Fig. 2). This was a major limitation, particularly since each layer of graphene oxide introduced an additional layer of amorphous material adding to background contrast. With limited electrical conductivity heavily dependent upon the degree of oxidization [62], it was also unclear as to how the insulating properties of bound oxide groups would influence the charging of vitreous samples.

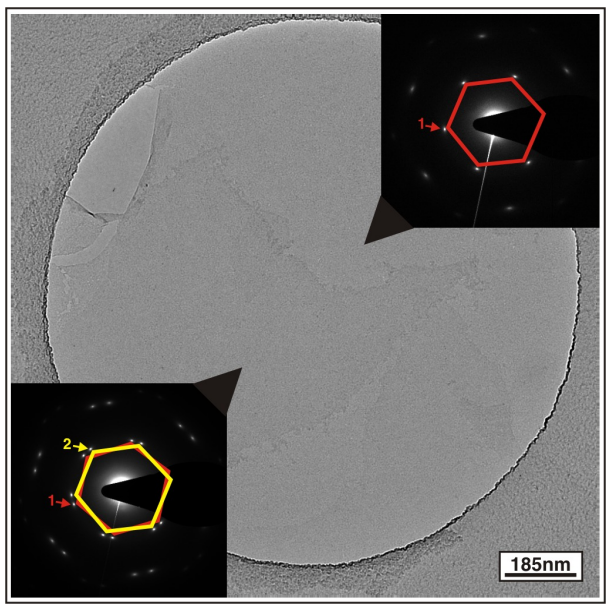

**Figure 2, The inadvertent stacking of graphene oxide layers:** An area of 1 µm perforated amorphous carbon spanned by single (a) and double (a) layers of freestanding graphene oxide (defocus -2 µm, pixel size 7.4 Å) as indicated by superimposed hexagonal diffraction patterns corresponding to each layer (insets).

The growth of continuous, large-area graphene by chemical vapour deposition (CVD) [63] addressed the fundamental limitation of graphene oxide supports and did not suffer the same attenuation of transparency or electrical conductivity. As expected, the power spectra from images of a clean, single-layer graphene demonstrate no phase contrast below its lattice resolution (Fig. 1, green; an image of clean single-layer graphene is shown e.g. in Fig. 6b) and the background amplitude is comparable to that of images from vacuum areas [38]. Early samples produced using Ni foils were more graphitic in nature given the high carbon solubility of Nickel [64]. Predominantly single-layer graphene pushed the fundamental limit of crystalline TEM supports and became feasible after the low carbon solubility and large grain size of Cu foils (<0.001 atom % at 1000°C as opposed to ~1.3 atom % [64]) was applied in a self-limiting CVD process producing mainly (> 95%) single-layer (~0.34 nm) continuous graphene [65]. A transfer-free method directly etched Cu foils after CVD growth to produce TEM grids with single-layer graphene spanning patterned 30-60 µm diameter holes [66]. However, the direct transfer of CVD graphene from Cu foils to perforated amorphous carbon supports provided freestanding areas of graphene and ample surrounding space with sufficient contrast for focusing away from the region of interest - as is required by cryo-EM [38]. After demonstrating the striking contrast of plasmid DNA across graphene without the necessity of metal shadowing [38], a subsequent study succeeded in arranging ordered arrays of DNA across graphene with the eventual aim of routinely identifying individual bases by HR-TEM [66].

However, the hydrophobic properties of graphene make the conventional preparation of biological samples infeasible, limiting the application of graphene in cryo-EM. Amorphous carbon supports are rendered hydrophilic by ion plasma, disrupting the surface and introducing (for example) –OH and C=O groups. However, graphene undergoes knock-on damage as incident ions sputter carbon from the basal plane. A doping method based on oxidative annealing was introduced, rendering graphene TEM supports sufficiently hydrophilic with minimal structural attenuation and maintained electrical conductivity [43]. Raman spectroscopy also indicated the removal of amorphous trace material during oxidative annealing (likely consumed as CO and $CO_2$ during oxidization) that in addition to indicative Raman shifts, further suggested oxidative doping rather than oxidation of sp-3 bound contaminants (i.e. the creation of something structurally closer to graphene oxide) [43].

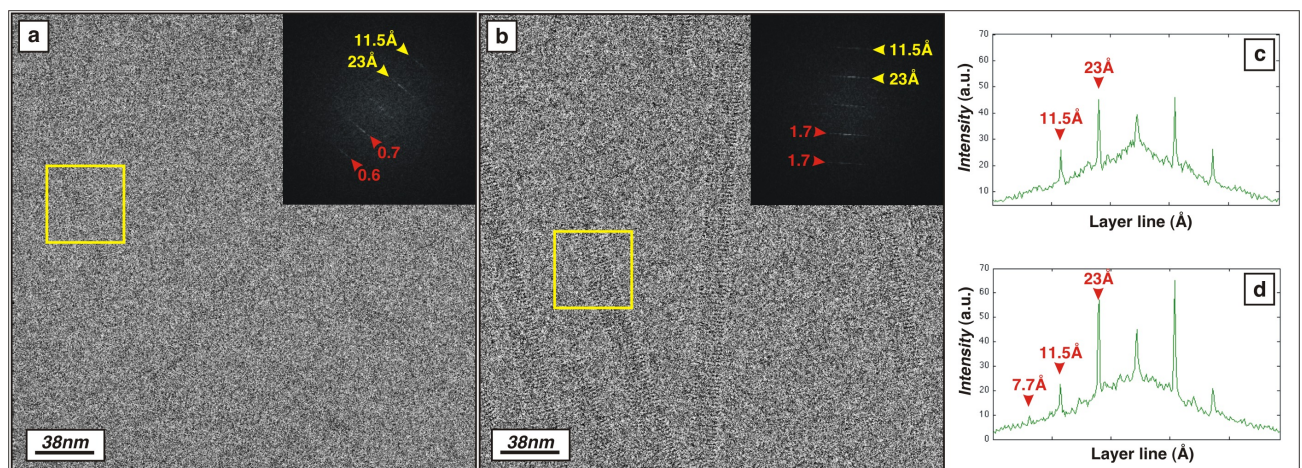

**Figure 3, The improved SNR of samples imaged across graphene as opposed to thin amorphous carbon:** TMV was imaged across freestanding areas of 2nm thick amorphous carbon (a) and single-layer graphene (b, c). The transparency of either support can be compared according to the SNR of the 3[rd] and 6[th] order layer lines (23 Å & 11.5 Å respectively) in the Fourier transform of individual TMV fibers (a & b insets, SNR's labeled in red). But what is most striking is the appearance of a 9[th] layer line (7.7 Å) in the FFT profile plot (c) sampled from another TMV/graphene micrograph (not shown) - attesting to the remarkable transparency and imaging properties of graphene.

With this method, the background of graphene and thin amorphous carbon supports were compared according to the signal-to-noise ratio (SNR) of layer lines diffracted by a periodic viral structure (Tobacco mosaic virus, TMV). On average the graphene samples demonstrated an increase in SNR of up to 100% [43]. Figure 3 shows an example of TMV prepared across graphene in which the SNR is improved by no less than 150% (Fig. 3b) as compared to 2nm amorphous carbon (Fig. 3a). What is perhaps most striking is the appearance of a 9th layer line at ~7.7Å, demonstrated by an FFT profile plot (Fig. 3c) calculated for another image (not shown) taken from the same published dataset [43]. It is difficult to quantitatively and definitively ascertain improved imaging stability afforded by the high electrical conductivity of graphene supports. However, a recent article compared vitrified preparations of TMV with the addition of nano-crystalline graphene and carbon nanotubes (CNT) [67]. In this case, results similar to those in Fig. 3c suggested that an improvement in signal compared to the freestanding vitrified sample (i.e. no additional graphene or CNT) was likely due to improved sample stability afforded by the graphene either by acting to dissipate sample charge, or by presenting a support structure of increased beam resistivity to reduce random physical specimen movements [10].

**Graphene supports in Materials science TEM**
There are some difficulties associated with the use of graphene supports in HR-TEM/STEM that must be overcome before their full potential may be exploited. Firstly, atomic-resolution images inevitably require high electron doses that may introduce knock-on damage across the support. Such radiation damage can be limited by limiting acceleration voltages (below a threshold of ~80-90 keV [68]). Figure 4 shows an area of graphene free from defects after an exposure of ~$10^{10}$ e-/nm[2] at 80keV (Fig. 4a). However, after and electron dose of only $10^7$ e-/nm[2] at 300keV, significant structural damage is apparent (Fig. 4b). This is not an issue when imaging biological molecules across graphene since the dose tolerance of the sample is several orders of magnitude lower (< 3000 $e^-/nm^2$ per exposure). However, atomic-resolution imaging of low-contrast adsorbates on graphene requires doses that prohibit operation beyond the knock-on threshold.

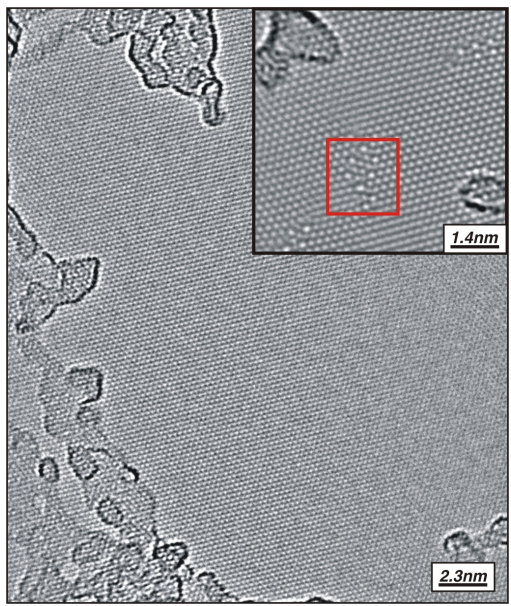

**Figure 4, Beam induced defects:** At 80keV the crystalline structure of graphene is stable at high doses but begins to demonstrate extensive defects left by knock-on damage (inset, red box) after comparatively low electron dose at 300keV (~$10^7$ $e^-/nm^2$). Note that the inset area was imaged at 80keV after brief dose at 300keV.

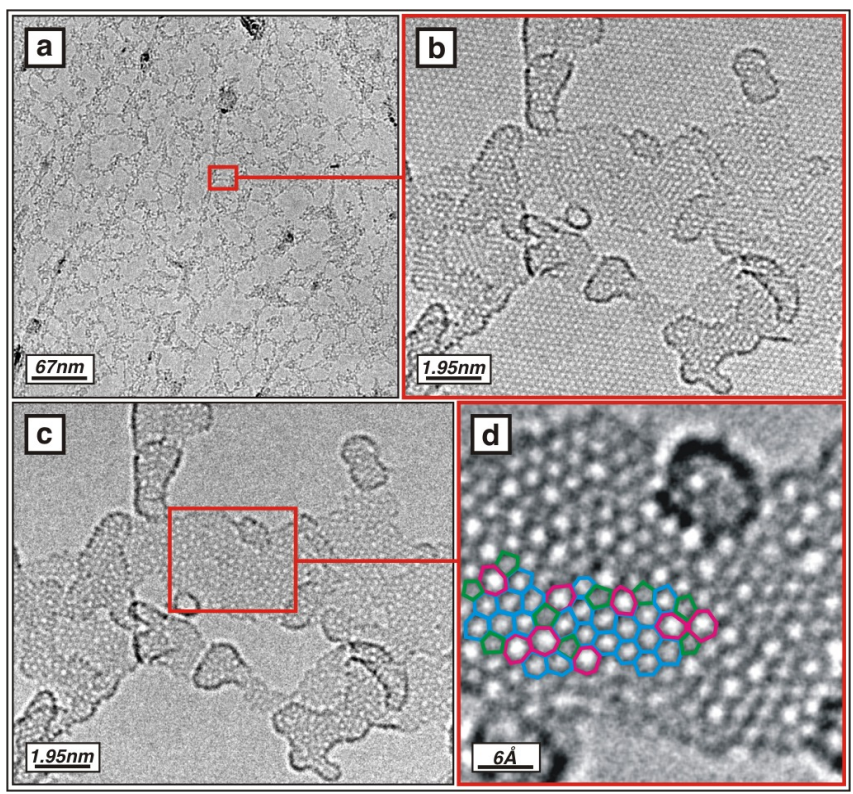

**Figure 5, HR-TEM images of contamination across graphene:** (a) Low magnification of a typical as-prepared graphene membrane. Featureless regions are atomically clean areas, while slightly darker grey patches are amorphous carbon contamination as shown in panel (b). (b) High-resolution TEM image of amorphous contamination, also after removing the graphene lattice by a Fourier filter (c). (d) Average of 8 frames, with structure indicated by geometric overlay. Overlay coloring corresponds to polygons of 5 (green), 6 (blue) and 7 (magenta) carbon atoms.

Graphene films prepared under ambient conditions contain significant amounts of contamination. Although significantly less than conventional amorphous carbon films, this contamination contributes to background signal. However, whilst this is an obstacle to atomic HR-TEM studies, in life sciences TEM the contribution of such adsorbates are perhaps no greater than that of buffer solution constituents in which samples are prepared and vitrified. Exclusive of these adsorbates, it is difficult to specifically attach objects to pristine graphene given the inert properties of the material (as in cryo-EM). Most likely, defect sites will be required to capture features of interest in a controlled manner.

Several early publications have discussed potential observations of light-element, contaminant molecules and atoms across graphene [48, 69-71]. Here we demonstrate the potential of graphene supports by studying slightly larger adsorbate "contamination" bound to the surface [72, 73]. This adsorbate is more stable under the beam; hence, its internal structure can be revealed. Figure 5 shows a low magnification image demonstrating the typical degree of contamination normally observable across graphene (Fig. 5a). We show an area of adsorbate spanning the graphene at atomic resolution before and after elimination of the graphene lattice from the image by Fourier filtering (Fig. 5b & c respectively). After filtering, the pristine graphene can no longer be distinguished (as if imaging vacuum), providing an unobstructed view of adsorbate's atomic structure (Fig. 5d). Although single exposures (Fig. 5b & c) yield much insight, the signal to noise ratio (SNR) is vastly improved after averaging several frames (Fig. 5d, averaging 8 frames). We clearly see the atomic structure (Fig. 5d, overlay) of the amorphous adsorbate across its thinnest area, resembling planar $sp^2$, yet amorphized graphene in its appearance and contrast [32]. This suggests the adsorbate is likely to consist primarily of stray carbon. Thus, the use of graphene supports at an acceleration voltage below the knock-on threshold facilitates the structural elucidation of thin, non-periodic (amorphous) adsorbates, as just demonstrated elsewhere [74].

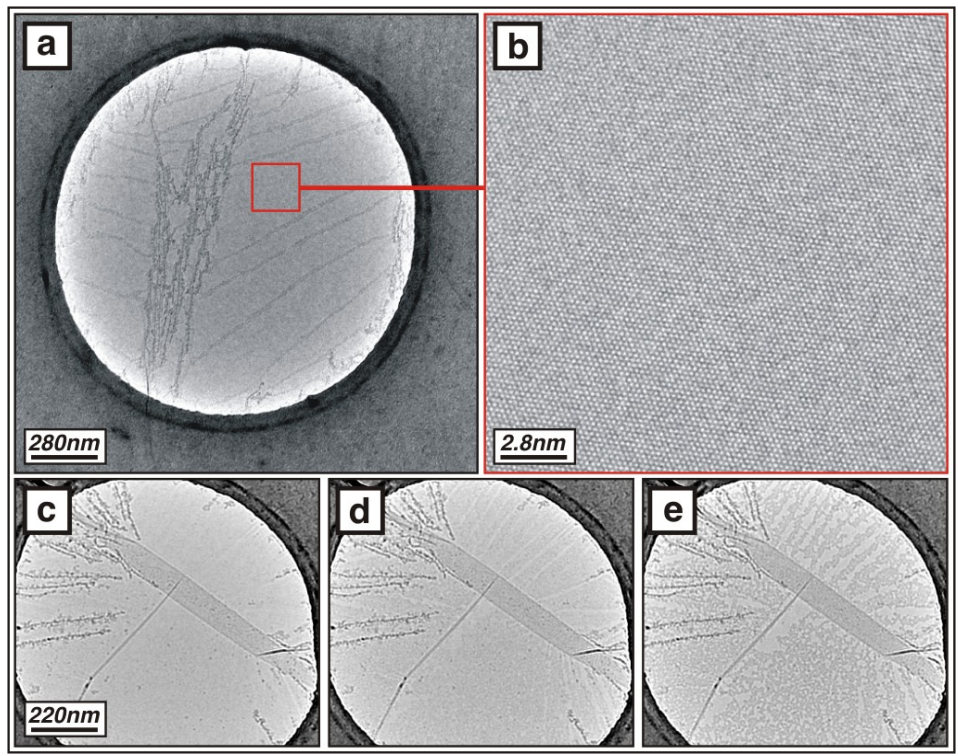

**Figure 6, The propagation of amorphous contamination:** (a) Pristine graphene after heating in vacuum (500°C, 1h). Note the amorphous decoration at grain boundaries and folds, otherwise the graphene is atomically pristine (b), within the delineated area in panel a). (c-e) Formation of amorphous contamination on the same sample after exposure to air (c) and subsequent dose of ~$10^4$ e$^-$/nm$^2$ (d) and ~$2*10^4$ e$^-$/nm$^2$ (d). (c-e) are at the same scale.

Having demonstrated the contamination found across graphene prepared in ambient conditions, the nature of this contamination is worth discussion. These membranes are prepared in atmosphere, either by mechanical cleavage or CVD synthesis, then transferred to TEM grids and heated (~200-300 °C) prior to insertion to the TEM. In all cases we see a landscape of pristine, crystalline areas interspersed with an amorphous carbon network (Fig. 5a & b). One way to remove this contamination is to heat the graphene membrane in the vacuum *prior* to beam exposure [45, 46, 75] (once the contamination is exposed to the beam, its fixation and transformation into amorphous carbon ensures it cannot easily be removed). In figure 6 we show a graphene membrane (monolayer, prepared by CVD across copper) that has been heated (before imaging) to 500°C in the TEM's vacuum for 1 hour. The graphene membrane is now largely pristine/atomically clean, with only grain boundaries [30] and folds [76] retaining contamination (Fig. 6a). After exposure of the same sample to ambient conditions (air) for ~10 minutes, subsequent observation reveals what are striking degrees of contamination after such brief exposure (Fig. 6c-e). Initially, the graphene is covered with a continuous film (Fig. 6c) that is quickly decomposed into the characteristic amorphous patches we often see intersecting areas of pristine graphene (Fig. 6e). Furthermore, the transition occurs at a rather low dose ($10^4 e^-/nm^2$, Fig. 6d). It is important to re-iterate that the adsorbate layer is present upon exposure, and changes morphology under the beam as opposed to deposition during irradiation (this is clearly demonstrated by the supplementary video). These results imply that ex-situ preparation and transfer through air cannot produce extended atomically clean graphene films. We may further conclude that the specific (clean) deposition of small objects across graphene will require (1) in-situ preparation of clean graphene in vacuum (e.g. heating) and (2), deposition of the "sample" (i.e. ad-atoms, molecules, clusters, etc.) in-situ or at least within the same vacuum system.

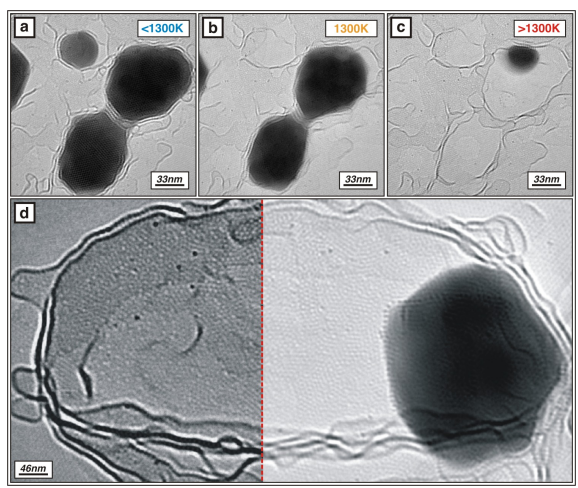

**Figure 7, Evaporation of gold particles by Joule heating across graphene:** (a-c) The gold particles melt during in-situ observation at temperatures of <1300K, 1300K and >1300K respectively, leaving behind an encapsulating carbon shell. (d) High-resolution image of the carbon shell. In the left part of the image, the underlying graphene lattice was removed by a Fourier filter and the contrast was increased. Adapted from [45, 46].

By passing electrical current, graphene TEM supports may also serve as an in-situ heating platform [45, 46, 77-79]. Joule heating of the graphene membrane can reach temperatures in excess of 2000 Kelvin [45, 46, 77, 78], at which carbon adsorbates reorganize into polycrystalline layers [46]. The transformation of gold particles (at melting point) can already be observed across graphene at lower temperatures. Figure 7 shows the heat-induced evaporation of gold particles across graphene as observed by HR-TEM (Fig. 7a-c). Whilst this phenomenon may also be studied using conventional amorphous carbon films, structural detail of an encapsulating carbon shell was previously difficult to obtain. Lee et al. first published images of an amorphous carbon shell encapsulating gold particles prepared across graphene, after using image post-processing (Fourier filtering) to remove the contrast of the gold particle [80]. Westenfelder et al. succeeded in removing the gold particles after heating, thereby isolating the carbon shells [45, 46]. As shown by Fig. 7d, the structure of the carbon shell is now apparent for further investigation. Importantly, we not only recognize the shape of the shell, but also remarkably observe a highly amorphous structure where layers of graphitic carbon appear in planar view. These results attest to the remarkable properties of graphene in observing miniscule detail consisting of as little as several disordered layers of graphitic carbon that would otherwise be impossible to analyze having used traditional amorphous carbon film.

**Conclusion**

Recent technical manuscripts have clearly demonstrated the benefits of graphene in the preparation of samples for cryo-EM - enhanced crystalline and electrical properties stand to drastically improve the stability and signal of weak-phase biological samples [38, 42-44, 59]. The highest resolution structure determined by cryo-EM to date is that of Aquaporin-0, solved using 2D electron crystallography to a resolution of 1.9Å [81]. Crucial limiting factors such as specimen flatness, stability and charging were addressed by sandwiching the 2D crystals between thin amorphous carbon films [13, 81]. A previous study has already demonstrated the potential benefits of conductive TiSi glass films in dissipating charge and reducing consequent drift at high-tilt [82]. Hence, functionalized CVD graphene should be a direct substitute capable of maximizing SNR, reducing charging and perhaps most interestingly, providing an "atomically flat" support for 2D crystals. Although not specifically related to TEM, a study evaluated the biocompatibility of CVD graphene with the culture of mouse hippocampal neurons [83]. The promotion of neuron sprouting and outgrowth across CVD graphene compliments techniques where samples are cultured directly across TEM grids [84]. Electrically conductive graphene supports would not only improve imaging stability but perhaps also provide exciting possibilities for electrically stimulated and time-resolved studies of neuron structure by cryo-EM.

Concerning materials science applications, the insights from recent, atomic-resolution images of adsorbates across graphene can be divided into three categories. First would be the imaging of "contamination" across graphene [48, 69, 71], including the example presented in Fig. 5. Importantly, these high-resolution, high-signal-to-noise ratio images of small clusters of light-element contamination, indicate that similar quality high-resolution images might be obtained from small molecular clusters [85] that will require controlled deposition across a clean graphene support. However, this relies upon addressing the experimental difficulties discussed previously. The second category demonstrates how graphene has proven itself essential to in-situ experiments [46, 78]. Here, the benefits of its high chemical inertness, mechanical and thermal stability as well as electrical conductivity, open new avenues for in-situ EM that go beyond reducing background signal. In-situ heating has also alleviated (to some extent) contamination issues and allowed new insights into low-contrast features such as adsorbed objects. The final category includes the straightforward use of graphene as a transparent sample support for high-resolution imaging. This has only been demonstrated in a few cases so far. Lee et al [80] demonstrated the visualization of the soft-hard interface, where the "soft" part would otherwise not be visible with conventional supports. Extremely small Co-based nanocrystals [86] and cBSA-Quantum dots [89] have also been imaged with remarkable SNR having been prepared across graphene.

Some other applications further demonstrate that aside from uniform crystalline background, a wider range of additionally outstanding material properties make graphene an interesting material for TEM investigation, and are worth noting. For example, Mohatny et al., wrapped bacteria in graphene to retain bound water during imaging [87]. Graphene has also been employed as an electron-transparent membrane for environmental cells [88].

The use of graphene as a support film is promising for a variety of microscopic applications, but this potential is not easily exploited. Amidst the rapid development of experimental techniques driven by a much larger range of potential applications, what is at this point essential is the large-scale availability of suitable graphene TEM supports and simplified handling/preparation methods exercisable in any lab largely exclusive from specialized equipment. In materials science the removal of unwanted adsorbates and the controlled deposition of target objects is a major obstacle. However, with the ongoing progress in graphene synthesis and device preparation it is likely that we will see TEM supports as one of graphene's primary applications.